# Force, Torque, Linear Momentum, and Angular Momentum in Classical Electrodynamics

Masud Mansuripur

College of Optical Sciences, The University of Arizona, Tucson, Arizona 85721

**Abstract**. The classical theory of electrodynamics is built upon Maxwell's equations and the concepts of electromagnetic (EM) field, force, energy, and momentum, which are intimately tied together by Poynting's theorem and by the Lorentz force law. Whereas Maxwell's equations relate the fields to their material sources, Poynting's theorem governs the flow of EM energy and its exchange between fields and material media, while the Lorentz law regulates the back-and-forth transfer of momentum between the media and the fields. An alternative force law, first proposed by Einstein and Laub, exists that is consistent with Maxwell's equations and complies with the conservation laws as well as with the requirements of special relativity. While the Lorentz law requires the introduction of hidden energy and hidden momentum in situations where an electric field acts on a magnetized medium, the Einstein-Laub (E-L) formulation of EM force and torque does not invoke hidden entities under such circumstances. Moreover, *total* force/torque exerted by EM fields on any given object turns out to be independent of whether the density of force/torque is evaluated using the law of Lorentz or that of Einstein and Laub. Hidden entities aside, the two formulations differ only in their predicted force and torque *distributions* inside matter. Such differences in distribution are occasionally measurable, and could serve as a guide in deciding which formulation, if either, corresponds to physical reality.

**1. Introduction**. In the mid 1960s, Shockley discovered a problem with the classical theory of electromagnetism. Under certain circumstances involving magnetic matter in the presence of an electric field, Shockley found that the momentum of the EM system is not conserved [1-3]. He attributed the momentum imbalance to a certain amount of "hidden momentum" residing inside magnetic media subjected to electric fields. In doing so, Shockley kept Maxwell's equations and the Lorentz force law from colliding with the universal law of momentum conservation. Subsequently, other authors elaborated on (and provided physical insight and justification for) the notion of hidden momentum [4-20].

Had Shockley used an alternative force law proposed in 1908 by Albert Einstein and Jacob Laub [21], he would have found that the combination of Maxwell's macroscopic equations and the Einstein-Laub (E-L) law complies with the conservation laws *without* the need for hidden entities. Moreover, he would have recognized that all known measurements of force and torque on rigid bodies supporting the Lorentz law also agree with the E-L theory. In other words, rather than introducing hidden entities into the electrodynamics of magnetic media, one could as well adopt the E-L force law without violating the experimentally established facts of physics. The present paper aims to place the E-L theory in a broader context, countering the suspicion that perhaps something untrustworthy lurks beneath the surface of this particular formulation of EM force and torque.

It must be emphasized at the outset that the advantages of the E-L formulation vis-a-vis hidden energy and hidden momentum are limited to those instances where hidden entities are invoked in conjunction with magnetic media. There exist legitimate uses for hidden entities outside the domain of magnetic materials as, for example, in the case of a spinning electrically-charged shell subjected to an external electric field, where "hidden" momentum is carried by the internal stresses of the shell material [22]. Similarly, the electro-magneto-static situations examined in [17], where no magnetic matter is present, provide excellent examples of the proper role of hidden entities within the theory of electromagnetism. Generally speaking, the E-L theory treats the exchanges of energy and momentum between EM fields and magnetic matter without the complication of hidden entities. In all other instances where the E-L law has the same expression as the Lorentz law (e.g., force and torque exerted by EM fields on free charges and free currents), one should not hesitate to invoke the hidden entities if and when the need arises.

Much has been made of the letter in which Einstein himself, in response to a June 15, 1918 letter from Walter Dällenbach concerning the EM stress-energy tensor, wrote: *"It has long been known that the values I had derived with Laub at the time are wrong; Abraham, in particular, was the one who presented this in a thorough paper. The correct strain tensor has incidentally already been pointed out by Minkowski"* [23]. We now know, however, that the major difference between the Lorentz and E-L formulations is the absence of hidden entities (within magnetic materials) in the latter. In other words, it can be shown that *total* force and *total* torque exerted by EM fields on any object are precisely the same in the two formulations, provided that the contributions of hidden momentum to the Lorentz force and torque exerted on magnetic matter are properly subtracted [24,25]. Since the vast majority of the experimental tests of the Lorentz law pertain to total force and/or total torque experienced by rigid bodies, these experiments can be said to equally validate the E-L formulation.

For the sake of completeness, Sec. 8 will briefly address the Abraham and Minkowski strain tensors mentioned in Einstein's letter to Dällenbach, highlighting their substantial differences not only with the E-L law, but also with the conventional Lorentz law.

**2. Synopsis**. Whereas Maxwell's equations are unique and undisputed, there exist alternative expressions for EM force, torque, energy, and momentum in the classical literature. The focus of the present paper is on two different approaches to the latter aspects of electrodynamics, one that can loosely be associated with the name of H. A. Lorentz, and another that could be traced to A. Einstein and J. Laub. While in the Lorentz approach, electric and magnetic dipoles are reduced to bound electrical charges and currents, the E-L treatment considers dipoles as independent entities, on a par with free electrical charges and currents. Section 3 argues that Maxwell's macroscopic equations permit different models (or interpretations) of the electric and magnetic dipoles to coexist. Different models lead to different versions of the Poynting theorem, as discussed in Sec. 4, and also to different expressions for the EM force and momentum densities, as elaborated in Sec. 5. It will be seen that the Lorentz and E-L formalisms, although differing in intermediate steps, generally yield similar results in the end.

The E-L expression of EM force-density may be parsed in different ways, assigning different contributions by the various constituents of matter (i.e., charge, current, polarization, and magnetization) to the overall force-density. This subject is taken up in Sec. 6, in the context of certain objections raised against the E-L approach. Another objection, involving the expression of EM torque in the E-L formalism, is addressed in Sec. 7. Here it will be shown that the general expression of EM torque-density must include the terms $\boldsymbol{P} \times \boldsymbol{E}$ and $\boldsymbol{M} \times \boldsymbol{H}$, which were mentioned only briefly in Einstein and Laub's original paper [21], but perhaps their generality has not been sufficiently appreciated. Section 8 is devoted to a comparison between the EM force densities derived from the Abraham and Minkowski tensors on the one hand, and those supported by the E-L and Lorentz theories on the other. The absence of electrostrictive and magnetostrictive contributions to the force-density in both the Abraham and Minkowski theories stands in sharp contrast to the presence of these effects in the E-L formulation.

**3. Maxwell's macroscopic equations**. We take Maxwell's macroscopic equations [26] as our point of departure. In addition to free charge and free current densities, $\rho_{\text{free}}(\boldsymbol{r},t)$ and $\boldsymbol{J}_{\text{free}}(\boldsymbol{r},t)$, the macroscopic equations contain polarization $\boldsymbol{P}(\boldsymbol{r},t)$ and magnetization $\boldsymbol{M}(\boldsymbol{r},t)$ as sources of the EM field. It is important to recognize that Maxwell's equations, taken at face value, do not make any assumptions about the nature of $\boldsymbol{P}$ and $\boldsymbol{M}$, nor about the constitutions of electric and magnetic dipoles. These equations simply take polarization and magnetization as they exist in Nature, and enable one to calculate the fields $\boldsymbol{E}(\boldsymbol{r},t)$ and $\boldsymbol{H}(\boldsymbol{r},t)$ whenever and wherever the spatio-temporal distributions of the sources ($\rho_{\text{free}}, \boldsymbol{J}_{\text{free}}, \boldsymbol{P}, \boldsymbol{M}$) are fully specified.

In Maxwell's macroscopic equations, $\boldsymbol{P}$ is combined with the $\boldsymbol{E}$ field, and $\boldsymbol{M}$ with the $\boldsymbol{H}$ field, which then appear as the displacement $\boldsymbol{D}(\boldsymbol{r},t) = \varepsilon_0 \boldsymbol{E} + \boldsymbol{P}$ and the magnetic induction $\boldsymbol{B}(\boldsymbol{r},t) = \mu_0 \boldsymbol{H} + \boldsymbol{M}$. In their most general form, Maxwell's macroscopic equations are written

$$\boldsymbol{\nabla} \cdot \boldsymbol{D} = \rho_{\text{free}}, \tag{1a}$$

$$\boldsymbol{\nabla} \times \boldsymbol{H} = \boldsymbol{J}_{\text{free}} + \partial \boldsymbol{D}/\partial t, \tag{1b}$$

$$\boldsymbol{\nabla} \times \boldsymbol{E} = -\partial \boldsymbol{B}/\partial t, \tag{1c}$$

$$\boldsymbol{\nabla} \cdot \boldsymbol{B} = 0. \tag{1d}$$

It is possible to interpret the above equations in different ways, without changing the results of calculations. In what follows, we rely on two different interpretations of the macroscopic equations. This is done by simply re-arranging the equations without changing their physical content. We shall refer to the two re-arrangements (and the corresponding interpretations) as the "Lorentz formalism" and the "Einstein-Laub formalism."

In the Lorentz formalism, Eqs.(1a) and (1b) are re-organized by eliminating the $\boldsymbol{D}$ and $\boldsymbol{H}$ fields. The re-arranged equations are subsequently written as

$$\varepsilon_0 \boldsymbol{\nabla} \cdot \boldsymbol{E} = \rho_{\text{free}} - \boldsymbol{\nabla} \cdot \boldsymbol{P}, \tag{2a}$$

$$\boldsymbol{\nabla} \times \boldsymbol{B} = \mu_0 \left(\boldsymbol{J}_{\text{free}} + \frac{\partial \boldsymbol{P}}{\partial t} + \mu_0^{-1} \boldsymbol{\nabla} \times \boldsymbol{M}\right) + \mu_0 \varepsilon_0 \frac{\partial \boldsymbol{E}}{\partial t}, \tag{2b}$$

$$\boldsymbol{\nabla} \times \boldsymbol{E} = -\frac{\partial \boldsymbol{B}}{\partial t}, \tag{2c}$$

$$\boldsymbol{\nabla} \cdot \boldsymbol{B} = 0. \tag{2d}$$

In this interpretation, electric dipoles appear as bound electric charge and bound electric current densities ($-\boldsymbol{\nabla} \cdot \boldsymbol{P}$ and $\partial \boldsymbol{P}/\partial t$), while magnetic dipoles behave as Amperian current loops with a bound current-density given by $\mu_0^{-1} \boldsymbol{\nabla} \times \boldsymbol{M}$. None of this says anything at all about the physical nature of the dipoles, and whether, in reality, electric dipoles are a pair of positive and negative electric charges joined by a short spring, or whether magnetic dipoles are small, stable loops of electrical current. All one can say is that eliminating $\boldsymbol{D}$ and $\boldsymbol{H}$ from Maxwell's equations has led to a particular form of these equations which is consistent with the above "interpretation" concerning the physical nature of the dipoles.

Next, consider an alternative arrangement of Maxwell's equations, one that may be designated as the departure point for the E-L formulation. Eliminating $\boldsymbol{D}$ and $\boldsymbol{B}$ from Eqs.(1), one arrives at

$$\varepsilon_o \boldsymbol{\nabla} \cdot \boldsymbol{E} = \rho_{\text{free}} - \boldsymbol{\nabla} \cdot \boldsymbol{P}, \tag{3a}$$

$$\boldsymbol{\nabla} \times \boldsymbol{H} = \left(\boldsymbol{J}_{\text{free}} + \frac{\partial \boldsymbol{P}}{\partial t}\right) + \varepsilon_o \frac{\partial \boldsymbol{E}}{\partial t}, \tag{3b}$$

$$\boldsymbol{\nabla} \times \boldsymbol{E} = -\frac{\partial \boldsymbol{M}}{\partial t} - \mu_o \frac{\partial \boldsymbol{H}}{\partial t}, \tag{3c}$$

$$\mu_o \boldsymbol{\nabla} \cdot \boldsymbol{H} = -\boldsymbol{\nabla} \cdot \boldsymbol{M}. \tag{3d}$$

In the E-L interpretation, the electric dipoles appear as a pair of positive and negative electric charges tied together by a short spring (exactly as in the Lorentz formalism). However, each magnetic dipole now behaves as a pair of north and south poles joined by a short spring. In other words, magnetism is no longer associated with an electric current density, but rather with bound magnetic charge and bound magnetic current densities, $-\boldsymbol{\nabla} \cdot \boldsymbol{M}$ and $\partial \boldsymbol{M}/\partial t$, respectively. We emphasize once again that such interpretations have nothing to do with the physical reality of the dipoles. The north and south poles mentioned above are not necessarily magnetic monopoles (i.e., in the sense of the Gilbert model [19]); rather, they are "fictitious" charges that acquire meaning only when Maxwell's equations are written in the form of Eqs.(3).

The two forms of Maxwell's equations given by Eqs.(2) and (3) are identical, in the sense that, given the source distributions $(\rho_{\text{free}}, \boldsymbol{J}_{\text{free}}, \boldsymbol{P}, \boldsymbol{M})$, these two sets of equations predict exactly the same EM fields $(\boldsymbol{E}, \boldsymbol{D}, \boldsymbol{B}, \boldsymbol{H})$ throughout space and time. How one chooses to "interpret" the physical nature of the dipoles is simply a matter of taste and personal preference. Such interpretations are totally irrelevant as far as the solutions of Maxwell's equations are concerned.

**4. Electromagnetic energy**. Different interpretations of Maxwell's macroscopic equations lead to different expressions for the EM energy-density and the energy flow-rate (i.e., the Poynting vector). However, as will be seen below, the end results turn out to be the same if hidden energy is properly taken into account.

In the Lorentz formulation, we dot-multiply $\boldsymbol{B}(\boldsymbol{r},t)$ into Eq.(2c), then subtract it from the dot-product of $\boldsymbol{E}(\boldsymbol{r},t)$ into Eq.(2b). Defining the Poynting vector (in the Lorentz formalism) as

$$\boldsymbol{S}_L = \mu_o^{-1} \boldsymbol{E} \times \boldsymbol{B}, \tag{4}$$

we arrive at

$$\boldsymbol{\nabla} \cdot \boldsymbol{S}_L + \frac{\partial}{\partial t}(\tfrac{1}{2}\varepsilon_o \boldsymbol{E} \cdot \boldsymbol{E} + \tfrac{1}{2}\mu_o^{-1} \boldsymbol{B} \cdot \boldsymbol{B}) + \boldsymbol{E} \cdot \left(\boldsymbol{J}_{\text{free}} + \frac{\partial \boldsymbol{P}}{\partial t} + \mu_o^{-1} \boldsymbol{\nabla} \times \boldsymbol{M}\right) = 0. \tag{5}$$

Thus, in the Lorentz interpretation, EM energy flows at a rate of $\boldsymbol{S}_L$ (per unit area per unit time), the stored energy-density in the $\boldsymbol{E}$ and $\boldsymbol{B}$ fields is

$$\mathcal{E}_L(\boldsymbol{r},t) = \tfrac{1}{2}\varepsilon_o \boldsymbol{E} \cdot \boldsymbol{E} + \tfrac{1}{2}\mu_o^{-1} \boldsymbol{B} \cdot \boldsymbol{B}, \tag{6}$$

and energy is exchanged between fields and matter at a rate of

$$\frac{\partial}{\partial t}\mathcal{E}_L^{(\text{exch})}(\boldsymbol{r},t) = \boldsymbol{E} \cdot \boldsymbol{J}_{\text{total}} \quad \text{(per unit volume per unit time)}, \tag{7}$$

where $\boldsymbol{J}_{\text{total}}$ is the sum of free and bound current densities; see Eq.(5). Note that the exchange of energy between the fields and the material media is a two-way street: When $\boldsymbol{E} \cdot \boldsymbol{J}$ is positive, energy leaves the field and enters the material, and when $\boldsymbol{E} \cdot \boldsymbol{J}$ is negative, energy flows in the opposite direction. All in all, we have imposed our own interpretation on the various terms appearing in Eq.(5), which is the mathematical expression of energy conservation. The validity of Eq.(5), however, being a direct and rigorous consequence of Maxwell's equations, is independent of any specific interpretation.

A similar treatment of EM energy-density and flow-rate can be carried out in the Einstein-Laub approach. This time, we dot-multiply Eq.(3c) into $H(r,t)$ and subtract the resulting equation from the dot-product of $E(r,t)$ into Eq.(3b). We find

$$\nabla \cdot (E \times H) + \frac{\partial}{\partial t}(\tfrac{1}{2}\varepsilon_o E \cdot E + \tfrac{1}{2}\mu_o H \cdot H) + \left(E \cdot J_{\text{free}} + E \cdot \frac{\partial P}{\partial t} + H \cdot \frac{\partial M}{\partial t}\right) = 0. \qquad (8)$$

Thus, in the E-L interpretation, the Poynting vector is

$$S_{EL} = E \times H, \qquad (9)$$

the stored energy-density in the $E$ and $H$ fields is

$$\mathcal{E}_{EL}(r,t) = \tfrac{1}{2}\varepsilon_o E \cdot E + \tfrac{1}{2}\mu_o H \cdot H, \qquad (10)$$

and energy is exchanged between fields and media at the rate of

$$\frac{\partial}{\partial t}\mathcal{E}_{EL}^{(\text{exch})}(r,t) = E \cdot J_{\text{free}} + E \cdot \frac{\partial P}{\partial t} + H \cdot \frac{\partial M}{\partial t} \quad \text{(per unit volume per unit time)}. \qquad (11)$$

Once again, energy conservation is guaranteed by Eq.(8), which is a direct and rigorous consequence of Maxwell's equations, irrespective of how one might interpret the various terms of the equation.

It is noteworthy that the commonly-used Poynting vector $S = E \times H$ [26-28] is the one derived in the E-L formalism. The Poynting vector $S_L = \mu_o^{-1} E \times B$ associated with the Lorentz interpretation (and preferred by some authors [29,30]) has been criticized on the grounds that it does not maintain the continuity of EM energy flux across the boundary between two adjacent media [27]. The simplest example is provided by a plane EM wave arriving from free space at the flat surface of a semi-infinite magnetic dielectric at normal incidence. The boundary conditions associated with Maxwell's equations dictate the continuity of the $E$ and $H$ components that are parallel to the surface of the medium. Thus, at the entrance facet, the flux of energy associated with $S_L$ exhibits a discontinuity whenever the tangential $B$ field happens to be discontinuous. Proponents of the Lorentz formalism do not dispute this fact, but invoke the existence of a hidden energy flux at the rate of $\mu_o^{-1} M \times E$ that accounts for the discrepancy [31]. Be it as it may, since the hidden energy flux is not an observable, one cannot be blamed for preferring the formalism that avoids the use of hidden entities.

**5. Electromagnetic force and momentum**. In the Lorentz formalism, all material media are represented by charge and current densities. Generalizing the Lorentz law $f = q(E + V \times B)$, which is the force exerted on a point-charge $q$ moving with velocity $V$ in the EM fields $E$ and $B$, the force-density that is compatible with the interpretation of Maxwell's equations in accordance with Eqs.(2) may be written as follows:

$$F_L(r,t) = (\rho_{\text{free}} - \nabla \cdot P)E + \left(J_{\text{free}} + \frac{\partial P}{\partial t} + \mu_o^{-1}\nabla \times M\right) \times B. \qquad (12)$$

Substitution for the total charge and current densities from Eqs.(2a) and (2b) into the above equation, followed by standard manipulations, yields

$$F_L(r,t) = \overleftrightarrow{\nabla} \cdot (\varepsilon_o EE) + \overleftrightarrow{\nabla} \cdot (\mu_o^{-1} BB) - \tfrac{1}{2}\nabla(\varepsilon_o E \cdot E + \mu_o^{-1} B \cdot B) - \partial(\varepsilon_o E \times B)/\partial t. \qquad (13)$$

With the aid of the identity tensor $\overleftrightarrow{I}$, Eq.(13) may be rewritten as

$$\overleftrightarrow{\nabla} \cdot \left[\tfrac{1}{2}(\varepsilon_o E \cdot E + \mu_o^{-1} B \cdot B)\overleftrightarrow{I} - \varepsilon_o EE - \mu_o^{-1} BB\right] + \frac{\partial}{\partial t}(\varepsilon_o E \times B) + F_L(r,t) = 0. \qquad (14)$$

The bracketed entity on the left-hand-side of Eq.(14) is the Maxwell stress tensor $\overleftrightarrow{\mathcal{T}}(r,t)$ [26]. Thus the EM momentum-density in the Lorentz formalism is $G(r,t) = \varepsilon_0 E \times B = S_L/c^2$ (sometimes referred to as the Livens momentum). According to Eq.(14), the EM momentum entering through the closed surface of a given volume is equal to the change in the EM momentum stored within that volume plus the mechanical momentum transferred to the material media located inside the volume. The Lorentz force density $F_L(r,t)$ is simply a measure of the rate of transfer of momentum from the fields to the material media (or vice versa).

In the E-L formalism, the force-density, which has contributions from the $E$ and $H$ fields acting on the sources ($\rho_{\text{free}}, J_{\text{free}}, P, M$), is written [21]

$$F_{EL}(r,t) = \rho_{\text{free}} E + J_{\text{free}} \times \mu_0 H + (P \cdot \nabla)E + \frac{\partial P}{\partial t} \times \mu_0 H + (M \cdot \nabla)H - \frac{\partial M}{\partial t} \times \varepsilon_0 E. \quad (15)$$

Substitution from Eqs.(3) into the above equation, followed by standard algebraic manipulations, yields

$$F_{EL}(r,t) = (\nabla \cdot D)E + (B \cdot \nabla)H - \tfrac{1}{2}\mu_0 \nabla(H \cdot H) - \partial(E \times H/c^2)/\partial t$$
$$+ (D \cdot \nabla)E - \tfrac{1}{2}\varepsilon_0 \nabla(E \cdot E) + (\nabla \cdot B)H. \quad (16)$$

With the aid of the identity tensor $\overleftrightarrow{I}$, Eq.(16) is rewritten as

$$\overleftrightarrow{\nabla} \cdot \left[ \tfrac{1}{2}(\varepsilon_0 E \cdot E + \mu_0 H \cdot H)\overleftrightarrow{I} - DE - BH \right] + \frac{\partial}{\partial t}(E \times H/c^2) + F_{EL}(r,t) = 0. \quad (17)$$

The bracketed entity on the left-hand-side of Eq.(17) is the E-L stress tensor $\overleftrightarrow{\mathcal{T}}_{EL}(r,t)$ [21]. Thus, according to Einstein and Laub, the EM momentum-density is $G(r,t) = E \times H/c^2 = S_{EL}/c^2$, which is commonly known as the Abraham momentum. In words, Eq.(17) states that the EM momentum entering through the closed surface of a given volume is equal to the change in the Abraham momentum stored within the volume plus the mechanical momentum transferred to the material media located inside the volume. The E-L force-density $F_{EL}(r,t)$ is simply a measure of the momentum transfer rate from the fields to the material media (or vice versa).

Note that the stress tensors of Lorentz and E-L, when evaluated in the free-space region surrounding an isolated object, are exactly the same. This means that, in steady-state situations where the enclosed EM momentum does not vary with time, the force exerted on an isolated object in accordance with the Lorentz law is precisely the same as that predicted by Einstein and Laub. Even in situations which depart from the steady-state, the actual force exerted on an isolated object should remain the same in the two formulations. Here the difference between the EM momentum densities of Lorentz ($\varepsilon_0 E \times B$) and Einstein-Laub ($E \times H/c^2$), namely, $\varepsilon_0 E \times M$, accounts only for the mechanical momentum that is hidden inside magnetic dipoles [10,19,24]. Since hidden momentum has no observable effects on the force and torque exerted on material bodies [19], the difference in the EM momenta in the two formulations cannot have any physical consequences.

It is remarkable that Einstein and Laub proposed their force-density formula, Eq.(15), nearly six decades before Shockley discovered the lack of momentum balance in certain EM systems containing magnetic materials [1]. The concept of hidden momentum proposed by Shockley accounts for the momentum imbalance in EM systems that acquire mechanical momentum at a rate that differs from that dictated by the exerted Lorentz force. Had Shockley used the E-L force instead, he would have found perfect balance and no need for hidden momentum.

**6. Alternative expressions of the Einstein-Laub equation**. The E-L formula describing the EM force-density acting on matter has been criticized on several grounds. In this section, we briefly address these concerns and, where possible, suggest remedies.

A persistent criticism has been that the current $J_{\text{free}}$ in Eq.(15) is acted upon by $\mu_o H$ rather than by $B$, whereas experiments such as those involving Lorentz electron microscopy [32-34], or the deflection of charged particles passing through permanent magnets [35,36], seem to support the $J_{\text{free}} \times B$ formula [20]. One way to respond to this criticism is to note that the E-L theory provides an expression only for the *total* force-density exerted on a material medium containing free charge, free current, polarization, and magnetization. This total force may be parsed in different ways to yield different expressions for the force-density acting on the individual components $\rho_{\text{free}}, J_{\text{free}}, P$ and $M$. Equation (15) may thus be rewritten as follows:

$$\boldsymbol{F}_{EL}(\boldsymbol{r},t) = \rho_{\text{free}}\boldsymbol{E} + \boldsymbol{J}_{\text{free}} \times \boldsymbol{B} + (\boldsymbol{P}\cdot\boldsymbol{\nabla})\boldsymbol{E} + (\partial\boldsymbol{P}/\partial t)\times\boldsymbol{B}$$
$$+(\boldsymbol{M}\cdot\boldsymbol{\nabla})\boldsymbol{H} + \boldsymbol{M}\times(\boldsymbol{\nabla}\times\boldsymbol{H}) - \partial(\varepsilon_o\boldsymbol{M}\times\boldsymbol{E})/\partial t. \qquad (18)$$

Note in the above equation that both $J_{\text{free}}$ and $\partial P/\partial t$ now interact with the *B*-field (rather than with the *H*-field), and that $P$ and $M$ do not appear to behave symmetrically in response to the $E$, $B$ and $H$ fields. Note also that the last term of Eq.(18), associated with the force experienced by $M$, simply removes the contribution of the hidden momentum $\varepsilon_o M \times E$.

The alternative expression of the E-L force-density in Eq.(18) thus answers the criticism as to whether $\mu_o H$ or $B$ should act on the carriers of electrical current. In free space, where $B = \mu_o H$, this question is moot, of course, but the passage of electrical current through magnetic media requires further attention to interactions between moving particles (which comprise the current) and the stationary particles (which give rise to magnetism). Thus in experiments involving charged particles traveling through magnetic media (e.g., anomalous Hall effect [37], Lorentz electron microscopy [32]), neither the Lorentz nor the E-L force (irrespective of the manner in which the latter has been parsed) should suffice to describe the behavior of the system. Rather, one must also take into account particle-particle scatterings that might involve interactions such as spin-orbit coupling and quantum-mechanical exchange [37,38].

Returning to Eq.(18), the symmetry between $P$ and $M$ may be restored if we rewrite the final expression as

$$\boldsymbol{F}_{EL}(\boldsymbol{r},t) = \rho_{\text{free}}\boldsymbol{E} + \boldsymbol{J}_{\text{free}} \times \boldsymbol{B} + [(\boldsymbol{P}\cdot\boldsymbol{\nabla})\boldsymbol{E} + \boldsymbol{P}\times(\boldsymbol{\nabla}\times\boldsymbol{E})]$$
$$+[(\boldsymbol{M}\cdot\boldsymbol{\nabla})\boldsymbol{H} + \boldsymbol{M}\times(\boldsymbol{\nabla}\times\boldsymbol{H})] + \partial(\boldsymbol{P}\times\boldsymbol{B} - \boldsymbol{M}\times\boldsymbol{D} - \boldsymbol{P}\times\boldsymbol{M})/\partial t. \qquad (19)$$

In static situations, the last term of Eq.(19) drops out, and the remaining terms, for the specific parsing chosen here, provide exact expressions for the force-density exerted on the various constituents of matter. Also, in linear systems driven by a monochromatic (i.e., single-frequency $\omega$) excitation, time-averaging over each oscillation period $\tau = 2\pi/\omega$ removes from Eq.(19) the contribution of the last term. If, in addition to linearity, the material media are further assumed to be isotropic and non-absorptive, we may write [26-28]

$$\boldsymbol{P}(\boldsymbol{r},t) = \varepsilon_o[\varepsilon(\boldsymbol{r},\omega) - 1]\boldsymbol{E}(\boldsymbol{r},t), \qquad (20a)$$
$$\boldsymbol{M}(\boldsymbol{r},t) = \mu_o[\mu(\boldsymbol{r},\omega) - 1]\boldsymbol{H}(\boldsymbol{r},t), \qquad (20b)$$

where $\varepsilon$ and $\mu$ are the (real-valued) permittivity and permeability of the media at the excitation frequency $\omega$. The time-averaged E-L force-density of Eq.(19) may then be written

$$\langle \boldsymbol{F}_{EL}(\boldsymbol{r},t)\rangle = \langle \rho_{\text{free}}\boldsymbol{E} + \boldsymbol{J}_{\text{free}} \times \boldsymbol{B}\rangle + \tfrac{1}{2}\varepsilon_o[\varepsilon(\boldsymbol{r},\omega) - 1]\boldsymbol{\nabla}\langle \boldsymbol{E} \cdot \boldsymbol{E}\rangle$$

$$+ \tfrac{1}{2}\mu_o[\mu(\boldsymbol{r},\omega) - 1]\boldsymbol{\nabla}\langle \boldsymbol{H} \cdot \boldsymbol{H}\rangle. \qquad (21)$$

The above equation, which also covers static situations ($\omega = 0$), contains electrostrictive as well as magnetostrictive terms that are proportional, respectively, to the local gradients of the $E$- and $H$-field intensities, $\langle \boldsymbol{E} \cdot \boldsymbol{E}\rangle$ and $\langle \boldsymbol{H} \cdot \boldsymbol{H}\rangle$. This brings out a second criticism of the E-L theory, which is the alleged inadequacy of the magnitude of the electrostrictive term of Eq.(21) in accounting for the Hakim-Higham experiment involving the force of static electric fields on liquid dielectrics [39]. The interpretation of the Hakim-Higham experiment begins with the Abraham and Minkowski force-density equations (see Sec.8), neither of which contains a contribution from electrostriction. A phenomenological term is then added to the Abraham and Minkowski equations to produce the so-called "Helmholtz force" [40,41], which incorporates the needed electrostrictive effect. (For a description of the Helmholtz force, see endnote 1.) Hakim and Higham conclude that the Helmholtz force provides a better fit to their experimental data than does the E-L force. Interpretation of such experiments, however, as pointed out by Brevik [41], requires careful attention to spurious effects, and, in any case, it is necessary to examine a much broader range of static as well as dynamic situations before settling on a microscopic theory of EM force and torque that has a firm basis in physical reality.

**7. Electromagnetic torque and angular momentum**. The torque and angular momentum densities in the Lorentz formulation may be determined by cross-multiplying the position vector $\boldsymbol{r}$ into Eq.(14). We will have

$$\boldsymbol{r} \times \boldsymbol{F}_L(\boldsymbol{r},t) + \tfrac{\partial}{\partial t}(\boldsymbol{r} \times \boldsymbol{S}_L/c^2) + \tfrac{\partial}{\partial x}\{\boldsymbol{r} \times [\tfrac{1}{2}(\varepsilon_o\boldsymbol{E}\cdot\boldsymbol{E} + \mu_o^{-1}\boldsymbol{B}\cdot\boldsymbol{B})\hat{\boldsymbol{x}} - \varepsilon_o E_x\boldsymbol{E} - \mu_o^{-1}B_x\boldsymbol{B}\,]\}$$

$$+ \tfrac{\partial}{\partial y}\{\boldsymbol{r} \times [\tfrac{1}{2}(\varepsilon_o\boldsymbol{E}\cdot\boldsymbol{E} + \mu_o^{-1}\boldsymbol{B}\cdot\boldsymbol{B})\hat{\boldsymbol{y}} - \varepsilon_o E_y\boldsymbol{E} - \mu_o^{-1}B_y\boldsymbol{B}\,]\}$$

$$+ \tfrac{\partial}{\partial z}\{\boldsymbol{r} \times [\tfrac{1}{2}(\varepsilon_o\boldsymbol{E}\cdot\boldsymbol{E} + \mu_o^{-1}\boldsymbol{B}\cdot\boldsymbol{B})\hat{\boldsymbol{z}} - \varepsilon_o E_z\boldsymbol{E} - \mu_o^{-1}B_z\boldsymbol{B}\,]\} = 0. \qquad (22)$$

The first term on the left-hand side of Eq.(22) is the Lorentz torque-density, while the second term provides the time-rate-of-change of the EM angular momentum density. In the remaining terms, we have moved $\boldsymbol{r} \times$ inside the differential operators, which is readily justified by simple differentiation of the resulting expressions. The last three terms in Eq.(22) form the divergence of a 2$^{nd}$ rank tensor, thus confirming the conservation of angular momentum. The EM torque and angular momentum densities in the Lorentz formulation are thus given by

$$\boldsymbol{T}_L(\boldsymbol{r},t) = \boldsymbol{r} \times \boldsymbol{F}_L(\boldsymbol{r},t). \qquad (23)$$

$$\boldsymbol{\mathcal{L}}_L(\boldsymbol{r},t) = \boldsymbol{r} \times \boldsymbol{S}_L/c^2. \qquad (24)$$

A similar procedure can be carried out within the Einstein-Laub theory, but the end result, somewhat unexpectedly, turns out to be different. After cross-multiplying both sides of Eq.(17) into $\boldsymbol{r}$, one recognizes that $\boldsymbol{r} \times$ cannot move inside the last three differential operators without introducing certain additional terms. Taking account of the fact that $\partial \boldsymbol{r}/\partial x = \hat{\boldsymbol{x}}$, $\partial \boldsymbol{r}/\partial y = \hat{\boldsymbol{y}}$, and $\partial \boldsymbol{r}/\partial z = \hat{\boldsymbol{z}}$, one arrives at

$$\boldsymbol{r} \times \boldsymbol{F}_{EL}(\boldsymbol{r},t) + \boldsymbol{P} \times \boldsymbol{E} + \boldsymbol{M} \times \boldsymbol{H} + \tfrac{\partial}{\partial t}(\boldsymbol{r} \times \boldsymbol{S}_{EL}/c^2)$$

$$+ \tfrac{\partial}{\partial x}\{\boldsymbol{r} \times [\tfrac{1}{2}(\varepsilon_o\boldsymbol{E}\cdot\boldsymbol{E} + \mu_o\boldsymbol{H}\cdot\boldsymbol{H})\hat{\boldsymbol{x}} - D_x\boldsymbol{E} - B_x\boldsymbol{H}]\}$$

$$+\frac{\partial}{\partial y}\{r \times [½(\varepsilon_o E \cdot E + \mu_o H \cdot H)\hat{y} - D_y E - B_y H]\}$$

$$+\frac{\partial}{\partial z}\{r \times [½(\varepsilon_o E \cdot E + \mu_o H \cdot H)\hat{z} - D_z E - B_z H]\} = 0. \quad (25)$$

Once again, the last three terms in Eq.(25) form the divergence of a 2$^{nd}$ rank tensor, thus confirming the conservation of angular momentum. The remaining terms of the equation then yield expressions for the EM torque and angular momentum densities, namely,

$$T_{EL}(r,t) = r \times F_{EL} + P \times E + M \times H. \quad (26)$$

$$\mathcal{L}_{EL}(r,t) = r \times S_{EL}/c^2. \quad (27)$$

In their original paper [21], Einstein and Laub mentioned the need for the inclusion of $P \times E$ and $M \times H$ terms in the torque equation only briefly and with specific reference to anisotropic bodies. Of course, in linear, isotropic, lossless media, where $P$ is parallel to $E$ and $M$ is parallel to $H$, both cross-products vanish. However, in more general circumstances, the torque expression *must* include these additional terms. The above derivation of Eq.(25) should make it clear that the expression of EM torque in Eq.(26) has general validity, and that $P \times E$ and $M \times H$ must always be added to $r \times F_{EL}$ if the angular momentum of a closed system is to be conserved.

As was the case with the EM force discussed in Sec.5, the *total* EM torque exerted on an isolated object always turns out to be the same in the Lorentz and E-L formulations; any differences between the two approaches can be reconciled by subtracting the contribution of the hidden angular momentum density, $r \times (\varepsilon_o E \times M)$, from the Lorentz torque [19].

**8. Force and momentum according to Minkowski and Abraham**. The stress tensors of Minkowski [42] and Abraham [43,44] are essentially identical (see endnote 2 for clarification):

$$\overleftrightarrow{\mathcal{T}}(r,t) = ½(D \cdot E + B \cdot H)\overleftrightarrow{I} - DE - BH. \quad (28)$$

What distinguishes Minkowski's theory from that of Abraham is the EM momentum-density $G(r,t)$, which is $D \times B$ in Minkowski's case, and $E \times H/c^2$ in the case of Abraham [41,45]. Applying the divergence operator to the above stress tensor and invoking Maxwell's macroscopic equations and well-known vector identities, we find

$$\overleftrightarrow{\nabla} \cdot \overleftrightarrow{\mathcal{T}}(r,t) = -\rho_{\text{free}}E - (P \cdot \nabla)E - (M \cdot \nabla)H - \varepsilon_o E \times \frac{\partial B}{\partial t}$$

$$+\mu_o H \times \left(J_{\text{free}} + \frac{\partial D}{\partial t}\right) + ½\nabla(P \cdot E + M \cdot H). \quad (29)$$

From this point on, Eq.(29) must be treated in different ways, depending on whether the goal is to derive the Abraham or the Minkowski force-density. In the Abraham case we have

$$\overleftrightarrow{\nabla} \cdot \overleftrightarrow{\mathcal{T}}(r,t) = -\left[\rho_{\text{free}}E + J_{\text{free}} \times \mu_o H + (P \cdot \nabla)E + \frac{\partial P}{\partial t} \times \mu_o H + (M \cdot \nabla)H\right.$$

$$\left. - \frac{\partial M}{\partial t} \times \varepsilon_o E - ½\nabla(P \cdot E + M \cdot H)\right] - \partial(E \times H/c^2)/\partial t. \quad (30)$$

The last term in the above equation is the time-derivative of the Abraham momentum-density, $G_A(r,t) = E \times H/c^2$. Therefore, the bracketed terms constitute the Abraham force-density. This force-density is seen to differ from that of Einstein and Laub, Eq.(15), only in the gradient term $½\nabla(P \cdot E + M \cdot H)$.

To arrive at the Minkowski force-density, we return to Eq.(29) and proceed with our development of $\overleftrightarrow{\nabla} \cdot \overleftrightarrow{\mathcal{T}}$, as follows:

$$\overleftrightarrow{\nabla} \cdot \overleftrightarrow{\mathcal{T}}(\bm{r},t) = -[\rho_{\text{free}}\bm{E} + \bm{J}_{\text{free}} \times \bm{B} + (\bm{P} \cdot \bm{\nabla})\bm{E} + \bm{P} \times (\bm{\nabla} \times \bm{E}) - \tfrac{1}{2}\bm{\nabla}(\bm{P} \cdot \bm{E})$$
$$+ (\bm{M} \cdot \bm{\nabla})\bm{H} + \bm{M} \times (\bm{\nabla} \times \bm{H}) - \tfrac{1}{2}\bm{\nabla}(\bm{M} \cdot \bm{H})] - \partial(\bm{D} \times \bm{B})/\partial t. \qquad (31)$$

Since the last term in Eq.(31) is the time-derivative of the Minkowski momentum-density, $\bm{G}_M(\bm{r},t) = \bm{D} \times \bm{B}$, the bracketed terms constitute the force-density expression according to Minkowski. Despite apparent differences between the end results of Eqs.(30) and (31), the two expressions, arrived at via different routes, must be identical. The Abraham and Minkowski force densities differ from each other only by $\partial(\bm{D} \times \bm{B} - \bm{E} \times \bm{H}/c^2)/\partial t$.

To further simplify the Abraham force-density derived in Eq.(30), we specialize to the case of linear, isotropic, lossless media under monochromatic excitation. All functions of space and time must now be written as $f(\bm{r},t) = \text{Re}[\tilde{f}(\bm{r})\exp(-i\omega t)]$, where two-function products are time-averaged in accordance with $\langle f(\bm{r},t)g(\bm{r},t)\rangle = \tfrac{1}{2}\text{Re}[\tilde{f}(\bm{r})\tilde{g}^*(\bm{r})]$. Polarization and magnetization are related to the $\bm{E}$ and $\bm{H}$ fields via

$$\widetilde{\bm{P}}(\bm{r}) = \varepsilon_{\text{o}}[\varepsilon(\bm{r},\omega) - 1]\widetilde{\bm{E}}(\bm{r}), \qquad (32\text{a})$$

$$\widetilde{\bm{M}}(\bm{r}) = \mu_{\text{o}}[\mu(\bm{r},\omega) - 1]\widetilde{\bm{H}}(\bm{r}), \qquad (32\text{b})$$

where $\varepsilon$ and $\mu$ are real-valued functions of $\bm{r}$ and $\omega$. Under these circumstances, the time-averaged Abraham force-density becomes

$$\langle \bm{F}_A(\bm{r},t)\rangle = \tfrac{1}{2}\text{Re}\left\{\tilde{\rho}_{\text{free}}\widetilde{\bm{E}}^* + \widetilde{\bm{J}}_{\text{free}} \times \widetilde{\bm{B}}^* - \tfrac{1}{2}\varepsilon_{\text{o}}(\bm{\nabla}\varepsilon)\big|\widetilde{\bm{E}}\big|^2 - \tfrac{1}{2}\mu_{\text{o}}(\bm{\nabla}\mu)\big|\widetilde{\bm{H}}\big|^2\right\}. \qquad (33)$$

In the case of the Minkowski force-density given by Eq.(31), further simplification is achieved by specializing to linear isotropic media. Monochromaticity and time-averaging in this case would not be necessary if we limit the discussion to lossless, non-dispersive media, where

$$\bm{P}(\bm{r},t) = \varepsilon_{\text{o}}[\varepsilon(\bm{r}) - 1]\bm{E}(\bm{r},t), \qquad (34\text{a})$$

$$\bm{M}(\bm{r},t) = \mu_{\text{o}}[\mu(\bm{r}) - 1]\bm{H}(\bm{r},t). \qquad (34\text{b})$$

Subsequently, the Minkowski force-density may be written

$$\bm{F}_M(\bm{r},t) = \rho_{\text{free}}\bm{E} + \bm{J}_{\text{free}} \times \bm{B} - \tfrac{1}{2}\varepsilon_{\text{o}}(\bm{\nabla}\varepsilon)(\bm{E} \cdot \bm{E}) - \tfrac{1}{2}\mu_{\text{o}}(\bm{\nabla}\mu)(\bm{H} \cdot \bm{H}). \qquad (35)$$

It is readily observed that, upon time-averaging, the Minkowski force-density of Eq.(35) becomes identical to the (already time-averaged) Abraham force-density in Eq.(33). This should not be surprising, considering that the Abraham and Minkowski force-densities differ only by $\partial(\bm{D} \times \bm{B} - \bm{E} \times \bm{H}/c^2)/\partial t$, whose time-average is zero for monochromatic excitations in linear media (even in the presence of loss and anisotropy).

In piecewise homogeneous media, where $\varepsilon$ and $\mu$ within individual pieces of material do not vary with $\bm{r}$, and assuming $\rho_{\text{free}} = 0$ and $\bm{J}_{\text{free}} = 0$, the Abraham and Minkowski force densities of Eqs.(33) and (35) produce forces only at the boundaries between adjacent regions, where $\varepsilon$ and $\mu$ change discontinuously. As mentioned in Sec.6, the Abraham and Minkowski force densities do not possess electrostrictive and magnetostrictive terms, that is, terms proportional to $\bm{\nabla}\langle\bm{E} \cdot \bm{E}\rangle$ and $\bm{\nabla}\langle\bm{H} \cdot \bm{H}\rangle$, respectively. The addition of a phenomenological term to these equations (to arrive at the Helmholtz force) has thus been deemed necessary in order to explain certain experimental observations [41]. In contrast, the E-L theory has a built-in mechanism for producing electrostriction and magnetostriction, which gives it an advantage not only over the Abraham and Minkowski theories, but also over the Lorentz formulation [46].

**9. Concluding Remarks**. In applications involving rigid (as opposed to deformable) media, the E-L method yields results that are identical to those obtained in the Lorentz formalism, albeit *without* the need for hidden energy and hidden momentum within magnetic materials – the hidden entities being inescapable companions of the Lorentz approach. It may thus appear that the choice between the Lorentz and E-L formulations is a matter of taste; those who feel comfortable with hidden entities may continue to use the Lorentz law, while others can resort to the E-L theory in order to avoid keeping track of hidden entities. This apparent equivalence, however, does not stand up to further scrutiny. Even after subtracting the hidden momentum contribution from the Lorentz force, the corresponding force-density *distribution* within an object turns out to be substantially different from that predicted by the E-L theory. Such differences should be measurable [45-49] and, in fact, the scant experimental evidence presently available seems to favor the E-L approach [46]. Considering that in recent years it has become possible to trap small droplets of various liquids by means of focused laser beams [50], it would be desirable to excite one or more whispering gallery modes inside such trapped droplets, then monitor the deformations of the droplet as a function of the incident laser power, the excited mode indices, and the refractive index of the droplet. Detailed numerical simulations would be necessary to predict the deformation of the droplet in accordance with each and every one of the proposed EM stress-energy tensors. A comparison between the measurement results and the theoretical calculations should then enable one to decide, once and for all, the stress-energy tensor that represents the physical reality.

In 1973, Ashkin and Dziedzic performed a remarkable experiment in which they focused a green laser beam ($\lambda_0 = 0.53\,\mu m$) onto the surface of pure water [51]. They observed a bulge on the surface, where the focused laser beam had entered. Subsequent analysis by Loudon [52,53] showed that compressive radiation forces beneath the surface tend to squeeze the liquid toward the optical axis, causing a surface bulge via the so-called "toothpaste tube" effect. In his analysis, Loudon used the E-L formulation; his findings are consistent with the results of our computer simulations [46], which indicate a compressive force pointing everywhere toward the optical axis of the incident laser beam. In contrast, theoretical as well as numerical analyses based on the Lorentz formulation [46,54] reveal the existence of both expansive and compressive forces in different regions beneath the surface, which effectively cancel each other out, thus ruling out the possibility of bulge formation on the water surface. The observations of Ashkin and Dziedzic in [51] thus provide a rare experimental evidence against the Lorentz formulation and in support of the E-L force-density expression.

One must not forget, however, that the Abraham and Minkowski force densities also predict a bulge similar to that observed in the experiment, although, in this case, electrostriction (i.e., the second component of the Helmholtz force) is required to account for the "squeeze" of the liquid needed for stability. This alternative explanation of the observed bulge, discussed at length in [41], is qualitatively similar to Loudon's analysis based on the E-L equation [52,53]. Either way, it is clear that the experimental results hint at a departure from the Lorentz formulation, suggesting the need for further analysis in order to pinpoint the correct form of the microscopic force equation – one that could accurately predict the measurable characteristics of the bulge in addition to explaining other relevant observations [41,55-59]. Of particular interest here would be a detailed quantitative analysis of the experimental data pertaining to the coupling of light to elastic waves in nonlinear optics [60-64]. Stimulated Brillouin scattering in optical fibers [65,66] is a good example of such phenomena which could yield valuable information about the elastic deformation of the fiber's core region in the presence of intense light pulses. Needless to say, detailed knowledge of the EM field distribution in conjunction with numerical simulations of the elastic vibrations for the extant EM stress-energy tensors would be needed to decide which tensor comes closest to predicting the experimental observations.

To the best of our knowledge, there exist no experimental data on the coupling of EM waves to transparent magnetic materials. (For the purpose of deciding among the various stress-energy tensors, transparency of the magnetic medium at the operating wavelength is essential, given that thermal expansion or contraction effects should not be allowed to mix with the mechanical effects of radiation.) Substantial differences exist among the predictions of the various stress-energy tensors when a magnetic medium experiences EM force and/or torque in consequence of its interactions with optical or microwave radiation. Aside from the issue of hidden momentum, the EM force-density *distribution* throughout a magnetic medium is very much dependent on the assumed EM stress-energy tensor. Whereas in transparent non-magnetic media the electric polarization (i.e., density of electric dipoles) alone is responsible for the force and torque experienced by the medium, in magnetic materials both electric and magnetic dipoles contribute to the local force-density exerted by the fields on the material object. As before, elastic deformations of the magnetic medium (be it of a soft and flexible nature, or of such rigidity as to produce a measurable elastic response to the applied EM force) must be monitored and compared with theoretical calculations.

Finally, it has been predicted that the radiation pressure on a submerged mirror inside a liquid has a strong dependence on the phase angle $\varphi$ of the mirror's Fresnel reflection coefficient $\rho \cong |\rho| \exp(\mathrm{i}\varphi)$ [67,68]. Different stress tensors predict different radiation pressures on a submerged mirror whose phase angle $\varphi$ departs substantially from the usual value of 180°. It will be of considerable value if radiation pressure measurements similar to those of Jones *et al* [69-71] could be carried out inside liquids of varying refractive indices using submerged multilayer stack dielectric mirrors whose phase angles $\varphi$ substantially deviate from 180°. As explained in [58,68,72-74], the results of such measurements will help to distinguish among the various stress-energy tensors.

**Endnotes**

1. The Helmholtz force-density associated with the action of the *E* field on a dielectric material of mass density $\rho$ and relative permittivity $\varepsilon$ is often written as follows [41]:

$$\boldsymbol{F}_H(\boldsymbol{r},t) = -\tfrac{1}{2}\varepsilon_\mathrm{o}(\boldsymbol{\nabla}\varepsilon)(\boldsymbol{E}\cdot\boldsymbol{E}) + \tfrac{1}{2}\varepsilon_\mathrm{o}\boldsymbol{\nabla}\left(\rho\frac{\partial\varepsilon}{\partial\rho}\boldsymbol{E}\cdot\boldsymbol{E}\right).$$

The first term on the right-hand-side of the above equation arises naturally from the stress tensors of Abraham and Minkowski, as discussed in Sec.8. The second term, which is associated with electrostriction, is derived phenomenologically, using arguments from the theories of elasticity and thermodynamics [40,75].

2. In the literature [41,45,47,55], Abraham's stress tensor is usually written as a symmetrized version of Minkowski's tensor, that is,

$$\overset{\leftrightarrow}{\mathcal{T}}_A(\boldsymbol{r},t) = \tfrac{1}{2}\big[(\boldsymbol{D}\cdot\boldsymbol{E} + \boldsymbol{B}\cdot\boldsymbol{H})\overset{\leftrightarrow}{\mathbf{I}} - (\boldsymbol{D}\boldsymbol{E} + \boldsymbol{E}\boldsymbol{D}) - (\boldsymbol{B}\boldsymbol{H} + \boldsymbol{H}\boldsymbol{B})\big].$$

Abraham's concerns, as well of those of his followers, were primarily with linear, isotropic media, namely, media for which $\boldsymbol{D} = \varepsilon_\mathrm{o}\varepsilon\boldsymbol{E}$ and $\boldsymbol{B} = \mu_\mathrm{o}\mu\boldsymbol{H}$. In such cases, since the stress tensor of Minkowski, given by Eq.(28), is already symmetric, the above act of symmetrization does not modify the tensor. In Abraham's own paper [43], the stress tensor is written explicitly only twice, in Eqs.(Va) and (56), and in both instances it is identical to Minkowski's (asymmetric) tensor. At several points in his papers [43,44], Abraham mentions the symmetry of his tensor, but it appears that he has the special case of linear, isotropic media in mind. The special symmetry that Abraham introduced into Minkowski's theory is, of course, that between the energy flow rate, $\boldsymbol{E} \times \boldsymbol{H}$, and the electromagnetic momentum density, $\boldsymbol{E} \times \boldsymbol{H}/c^2$, which reside, respectively, in the fourth column and the fourth row of the stress-energy tensor. Be it as it may, in Eq. (28) we have chosen the asymmetric version of Abraham's (3×3) stress tensor, as it simplifies the subsequent discussion. In any event, this does not affect the main results and the conclusions reached in Sec. 8, since the media chosen for analysis in that section are linear and isotropic.